\documentclass[reprint,amsmath,amssymb,aps,superscriptaddress,showpacs]{revtex4-1}
\usepackage{graphicx}
\usepackage{mathrsfs}
\usepackage{bm}
\usepackage{amsmath}
\usepackage{dcolumn}
\usepackage{epstopdf}
\usepackage{dsfont}
\usepackage{amssymb}
\usepackage{tabularx}
\usepackage{array}
\usepackage{float}
\usepackage{color}
\usepackage{epstopdf}
\usepackage{mathrsfs}

\usepackage[colorlinks=true, letterpaper=true, pdfstartview=FitV, linkcolor=blue, citecolor=blue, urlcolor=blue]{hyperref}
\usepackage{color}
\begin{document}
\title{First-principles study of two-dimensional transition metal carbide M$_{\mathrm{n}+1}$C$_\mathrm{n}$O$_2$ (M=Nb,Ta)}


\author{Yue Shang}
\author{Yifan Han}
\author{Wenhui Wan}
\author{Yong Liu}
\author{Yanfeng Ge}
\email{yfge@ysu.edu.cn}\affiliation{Key Laboratory for Microstructural Material Physics of Hebei Province, School of Science, Yanshan University, Qinhuangdao, 066004, China}

\begin{abstract}

In the present work, the three stable MXenes M$_{\mathrm{n}+1}$C$_\mathrm{n}$O$_2$ (M=Nb,Ta) are explored based on first-principles calculations. These materials are important derivatives of 2D materials and exhibit distinctive properties, holding vast potential in nanodevices. All these M$_{\mathrm{n}+1}$C$_\mathrm{n}$O$_2$ (M=Nb,Ta) materials exhibit outstanding superconducting performance, with corresponding superconducting transition temperatures of 23.00K, 25.00K, and 29.00K. Analysis reveals that the high superconducting transition temperatures of MXenes M$_{\mathrm{n}+1}$C$_\mathrm{n}$O$_2$ (M=Nb,Ta) are closely associated with the high value of the logarithmic average of phonon frequencies, $\omega_{\log}$, and the strong electron-phonon coupling (EPC), attributed to the crucial contribution of low-frequency phonons. Additionally, we applied strain treatments of 2\% and 4\% to M$_{\mathrm{n}+1}$C$_\mathrm{n}$O$_2$ (M=Nb,Ta), resulting in varying changes in superconducting transition temperatures under different strains.

  \end{abstract}

\maketitle
\section{\label{sec:level1}INTRODUCTION}

Since 2004 when there was a groundbreaking discovery on graphene, the one-atom-thick layer of carbon atoms arranged in a hexagonal pattern has revolutionized the study of two-dimensional (2D) materials. With remarkable characteristics, including an elevated surface-area-to-volume ratio, exceptional tensile strength, and superior electrical conductivity, graphene is exceptionally suited for a diverse array of applications~\cite{{1},{2},{3},{4},{5}}. This pivotal advancement has instigated a surge in research efforts directed toward exploring other 2D materials, such as transition metal dichalcogenides (TMDCs), hexagonal boron nitride (h-BN), and phosphorene. These materials present a spectrum of electronic, optical, and mechanical properties, promising prospects for applications spanning electronics, optoelectronics, and energy storage~\cite{{6},{7},{8},{9},{10},{11},{12}}. The discovery of superconductivity within these materials not only unlocks new avenues in physics but also facilitates the exploration of superconductivity within a 2D framework. These researches shed light on intricate phenomena, such as electron pairing and the interplay between dimensionality and superconductivity~\cite{{13},{14},{15},{16},{17},{18},{19},{20},{21},{22},{23},{24},{25},{26}}. The ultra-thin nature of 2D superconductors endows them with a heightened susceptibility to external stimuli, such as magnetic and electric fields, thereby paving the way for innovative strategies to manipulate their properties. The emergence of 2D superconductors heralds the potential for groundbreaking technologies, including compact magnetic resonance imaging devices and precise magnetic field detectors tailored for advanced applications, such as single-spin detection. Illustrative instances of superconductivity within 2D materials include graphene, which achieves superconductivity through doping or specialized layering techniques. There are many significant studies documenting superconducting behaviors in materials like FeSe on SrTiO$_3$ substrates and NbSe$_2$, underscoring the versatility and potential of 2D superconductors~\cite{{27},{28},{29}}. Continuous research endeavors are imperative to fully harness the potential of 2D superconductors for future technological innovations.

2D MXenes, formed by selective etching of the MAX phase~\cite{{30},{31},{32},{33},{34}}, have garnered significant interest in recent years for their distinctive properties and wide-ranging applications. With diverse compositions comprising transition metals (M), carbon or nitrogen (X), and surface terminations (T), MXenes (M$_{\mathrm{n}+1}$X$_\mathrm{n}$T$_\mathrm{z}$) manifest a plethora of structures, facilitating exploration into its various properties such as excellent electrical conductivity, robust mechanical strength, hydrophilicity, and substantial specific surface area. MXenes find utility across domains including energy storage, electromagnetic interference shielding, sensing technology, flexible electronics, and catalysts for water splitting and other reactions. A noteworthy breakthrough by Talapin et al. involved surface group modification of Nb$_2$CCl$_2$ through substitution and elimination reactions in molten inorganic salts, resulting in a superconducting material with a critical temperature (T$_c$) of 7.1K~\cite{35} This breakthrough has expanded further possibilities for MXene research, particularly in the realm of superconductivity~\cite{{35},{36},{37},{38},{39},{40}}. Bekaert et al. identified six superconducting MXenes, where Mo$_2$N exhibited a maximum T$_c$ of 16K, indicating a successful synthesis in the experiment~\cite{{36},{37}}. Additionally, they examined the promotional effect of hydrogen passivation on superconductivity in Mo$_2$N and W$_2$N~\cite{38}. The discovery of superconductivity in MXenes has expanded their potential applications and is poised to propel further advancements in the field~\cite{39}. Continued research endeavors may unveil new MXene structures and properties, contributing to the development of innovative superconducting materials and technologies.

The properties of MXenes are amenable to manipulation through the modification of their surface terminations. Surface oxidation, leading to the formation of oxide layers, can modulate the surface chemistry, and influence the electronic structures of MXenes. In this study, we employ first-principles calculations to predict the superconductivity in 2D transition metal-carbon compounds M$_{\mathrm{n}+1}$C$_\mathrm{n}$O$_2$ (M=Nb,Ta).

\section{\label{sec:level1} COMPUTATIONS DETAILS}

First-principles calculations are conducted based on density functional theory (DFT), utilizing plane waves, and pseudopotentials within the Quantum ESPRESSO software package~\cite{41}.To mitigate interlayer interactions, M$_{\mathrm{n}+1}$C$_\mathrm{n}$O$_2$ (M=Nb,Ta) is simulated with a 15{\AA} vacuum layer in Materials Studio. Besides, we also employed the local density approximation method (LDA) to handle the electron exchange-correlation energy functionally, while utilizing norm-conserving pseudopotentials (NCPPs)~\cite{42} for numerical band structure calculations. In this study, a plane wave kinetic energy cutoff of 50 Ry is selected, and Brillouin zone integration is performed on a ${12}\times{12}\times{1}$ k-point grid for geometry optimization and density of states calculations. Furthermore, phonon dispersion calculations are executed on a ${6}\times{6 }\times{1}$ q-point grid. The EPW~\cite{43} method is also employed based on DFT, utilizing Wannier interpolation~\cite{44} (projecting s, p, and d orbitals of Nb (Ta) atoms and s and p orbitals of C (O) atoms) to accurately interpolate the band energies, different observables, and operator matrix. To ascertain the maximum localized Wannier functions results, a ${36}\times{36 }\times{1}$ k-point grid is set up.

The superconducting transition temperature T$_c$ is determined using the McMillan-Allen-Dynes formula~\cite{{45},{46}}

\begin{eqnarray}\label{COMPUTATIONAL}
\begin{split}
T_c =f_{1}f_{2}\dfrac{\omega_{log}}{1.2}\exp[\dfrac{-1.04(1+\lambda)}{\lambda-\mu^*_c(1+0.62\lambda)}]
 \end{split}
\end{eqnarray} 

where $\omega_{\log}$ represents the logarithmic average of the phonon frequencies, ${\mu_c}*$ denotes the effective screened Coulomb repulsion constant, empirically set within the range of 0.1-0.15

The logarithmic average $\omega_{log}$ is defined as:

\begin{eqnarray}\label{COMPUTATIONAL}
\begin{split}
\omega_{log}=\exp[\dfrac{2}{\lambda}\int{\dfrac{d\omega}{\omega}\alpha^2} F(\omega)ln(\omega)]
 \end{split}
\end{eqnarray}

Here, $f_1$ and $f_2$ represent the strong coupling and shape corrections, respectively:

\begin{eqnarray}\label{COMPUTATIONAL}
\begin{split}
f_{1}=\{1+[\dfrac{\lambda}{2.46(1+3.8\mu^*_c)}]^{3/2}\}^{1/3}
 \end{split}
\end{eqnarray} 

\begin{eqnarray}\label{COMPUTATIONAL}
\begin{split}
f_{2}=1+\dfrac{[(\omega_2/\omega_{log})-1]\lambda^2}{\lambda^2+3.312(1+6.3\mu^*_c)^2(\omega_2/\omega_{log})^2}
 \end{split}
\end{eqnarray} 

In this formula, $\omega_2$ is defined as: 

\begin{eqnarray}\label{COMPUTATIONAL}
\begin{split}
\omega_2=[\dfrac{2}{\lambda}\int{\alpha}^2 F(\omega)\omega d\omega]^{1/2}
 \end{split}
\end{eqnarray}

The total electron-phonon coupling (EPC) constant $\lambda$ can be obtained by summing the EPC constant $\lambda_{q\upsilon}$ for all phonon modes in the entire Brillouin zone (BZ) or by integrating the Eliashberg spectral function ${\alpha}^2 F(\omega)$ as follows:

\begin{eqnarray}\label{COMPUTATIONAL}
\lambda =\sum_{q\upsilon}\lambda_{q\upsilon}=2\int{\dfrac{{\alpha}^2 F(\omega)}{\omega}}d\omega
\end{eqnarray} 

\begin{eqnarray}\label{COMPUTATIONAL}
{\alpha}^2 F(\omega) =\dfrac{1}{2\pi N_F}\sum_{q\upsilon}\delta(\omega-\omega_{q\upsilon})\dfrac{\gamma_{q\upsilon}}{\hbar\omega_{q\upsilon}}
\end{eqnarray} 

In this formula, $E_F$ represents the state density at the Fermi level, $\omega_{q\upsilon}$ and $\gamma_{q\upsilon}$ denote the frequency and linewidth for phonon mode $\upsilon$ at wave vector q.

\begin{figure}[t!hp]
\centerline{\includegraphics[width=0.5\textwidth]{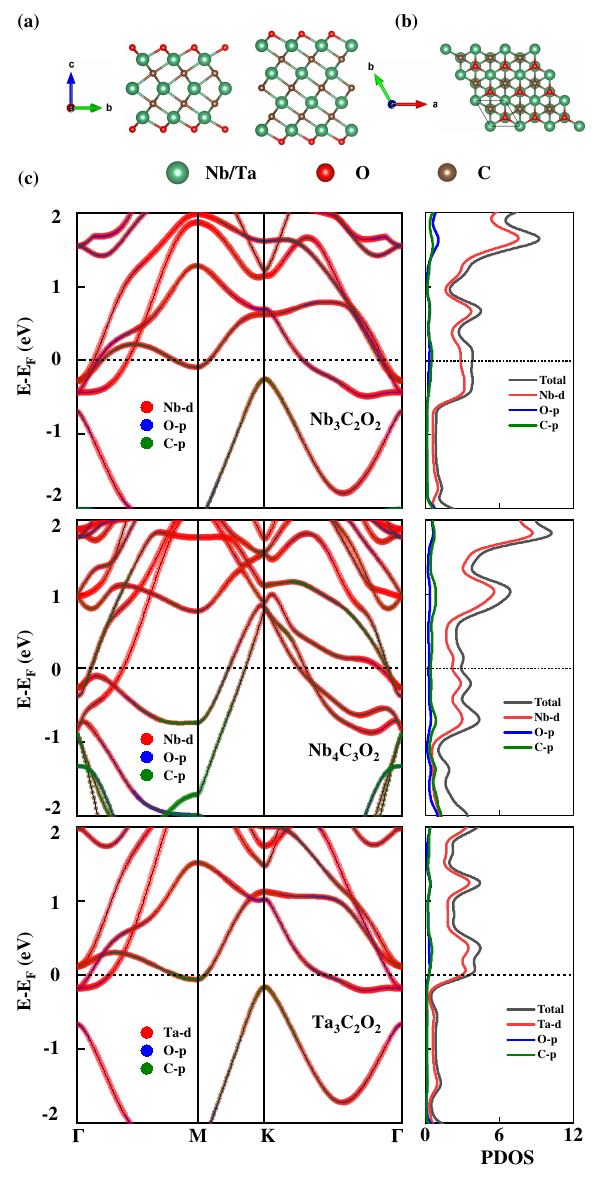}}
 \centering
\caption{Side view (a) and top view (b) of MXenes M$_{\mathrm{n}+1}$C$_\mathrm{n}$O$_2$ (M=Nb,Ta).(c) Band structures and density of states (DOS) of MXenes M$_{\mathrm{n}+1}$C$_\mathrm{n}$O$_2$ (M=Nb,Ta).
\label{fig:stru}}
\end{figure} 

\section{ RESULTS and DISCUSSION}

Figures 1(a) and 1(b) depict the atomic configurations of single-layer MXenes, M$_{\mathrm{n}+1}$C$_\mathrm{n}$O$_2$, where M can be either Nb or Ta. These configurations are represented alongside their specific symmetries: $D_{3h}$ for M$_3$C$_2$O$_2$ and $D_{3d}$ for Nb$_4$C$_3$O$_2$. In M$_3$C$_2$O$_2$, Nb or Ta atoms and C atoms are symmetrically arranged, where C atoms are positioned surrounding a central M atom. The Nb$_4$C$_3$O$_2$ structure organizes Nb and C atoms in a grid pattern, with oxygen atoms forming the outermost layers in both configurations. After detailed structural refinement, the lattice constant was established at 3.14{\AA}. Figure 1(c) delineates the electronic properties, revealing metallic characteristics characterized by band crossings at the Fermi level along the $\Gamma$-M-K-$\Gamma$ path for both Nb and Ta versions. Due to their shared group VB classification and equal valence electron counts, the band structures of Nb$_3$C$_2$O$_2$ and Ta$_3$C$_2$O$_2$ exhibit similarities, despite a discernible Fermi level shift from Nb to Ta. Notably, the analysis primarily focuses on the d orbitals of Nb (or Ta), with limited involvement from other atoms, categorizing these d orbitals into five subcategories. It is observed that the out-of-plane orbitals (d$z^2$, dxz, dyz) exhibit stronger interactions with the Fermi-level vibrations along the z-axis compared to the in-plane orbitals (d($x^2$-$y^2$), dxy). With further investigation into Table S1 data, we can observe a progressive increase in the density of states (DOS) at the Fermi level across Nb$_3$C$_2$, Nb$_3$C$_2$O$_2$, Nb$_4$C$_3$O$_2$, and Ta$_3$C$_2$O$_2$. This trend suggests that modifications such as oxygen incorporation, adjustments in carbon and niobium content, and the substitution of M atoms potentially enhance their superconducting properties rather than impede them.

\begin{figure}[tbp!]
\centerline{\includegraphics[width=0.5\textwidth]{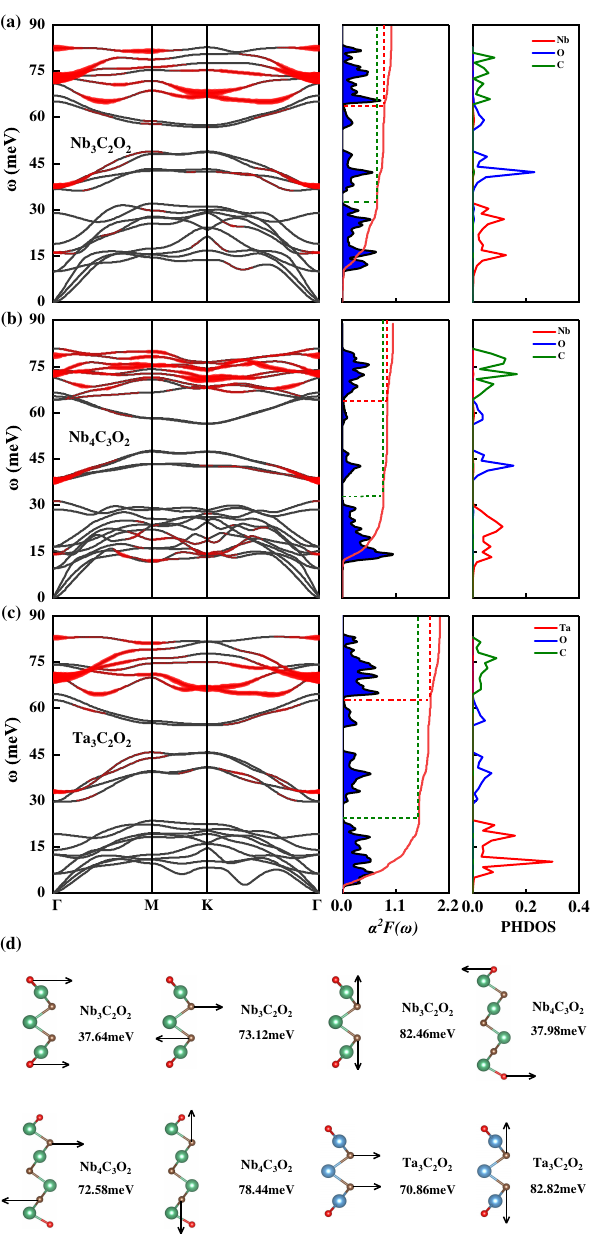}}
\caption{Phonon spectrum, Eliashberg spectral function (${\alpha}^2F(\omega)$), electron-phonon coupling constant ($\lambda$), phonon density of states (PHDOS) of MXenes (a)Nb$_3$C$_2$O$_2$, (b)Nb$_4$C$_3$O$_2$ and (c)Ta$_3$C$_2$O$_2$. (d) Vibration patterns of the optical phonon modes at the $\Gamma$ point.
\label{fig:E-V}}
\end{figure}

Analysis of Figure 2 data can provide pivotal insights into the phonon spectrum, Eliashberg spectral function (${\alpha}^2F(\omega)$), and EPC constant ($\lambda$) of MXenes M$_{\mathrm{n}+1}$C$_\mathrm{n}$O$_2$, with M being either Nb or Ta. One of the notable findings is the absence of imaginary frequencies within their Brillouin zones, which confirms the dynamical stability of these compounds. The phonon spectrum is segmented into two main regions: the low-frequency region (below $^{\sim}30$meV), predominantly characterized by Nb atom vibrations, and the mid-to-high-frequency region (above $^{\sim}30$meV), where vibrations are predominantly attributed to O and C atoms. Moreover, the study highlights the fact that EPC is chiefly influenced by low-energy phonons related to Nb or Ta atom vibrations. In Nb$_3$C$_2$O$_2$, Nb$_4$C$_3$O$_2$, and Ta$_3$C$_2$O$_2$, there are significant portions of $\lambda$ values originating from these metal atom vibrations, underscoring their crucial role in electron-phonon interactions. There is also an increase in $\lambda$ values with the inclusion of oxygen atoms, which notably enhances EPC. The enhancement is particularly notable for Nb and C atoms, showcasing significant improvements. Intriguingly, the substitution of Nb with Ta atoms leads to a considerable increase in EPC, underscoring the superior effectiveness of Ta in this regard. Furthermore, analysis of phonon linewidth at the $\Gamma$ point reveals distinct optical phonon vibration modes associated with O and C atoms in Nb$_3$C$_2$O$_2$ and Nb$_4$C$_3$O$_2$, whereas in Ta$_3$C$_2$O$_2$, only C atom vibrations are prominent. This analysis deepens our understanding of vibrational characteristics and electron-phonon interactions within these MXene structures.

The transverse acoustic (TA) mode, as illustrated in Figure S7, demonstrates a consistent softening trend moving from Nb$_3$C$_2$ to Nb$_3$C$_2$O$_2$ (and Nb$_4$C$_3$O$_2$), and finally to Ta$_3$C$_2$O$_2$. This underscores the significant roles that oxygen atoms and M atoms (where M equals Ta) play in promoting the softening of phonon modes. Such softening directly correlates with an enhancement in the EPC strength ($\lambda$), as observed in Figure S7(b). When we delve into the trends in Figures S7(a) and S7(b), we discovered that while the superconducting transition temperature (T$_c$) generally exhibits an upward trend, there is a noticeable dip in the transition from Nb$_3$C$_2$O$_2$ to Nb$_4$C$_3$O$_2$. Moreover, the increase in temperature from Nb$_3$C$_2$ to Nb$_3$C$_2$O$_2$ is substantially more pronounced than that from Nb$_3$C$_2$O$_2$ to Ta$_3$C$_2$O$_2$.

As in the McMillan-Allen-Dynes formula, the superconducting transition temperature (T$_c$) is determined by two critical parameters: the logarithmic average phonon frequency ($\omega_{log}$) and the EPC strength ($\lambda$). Higher values of $\omega_{log}$ and $\lambda$ generally lead to higher T$_c$ values. Analyzing the $\omega_{log}$ values from Table S1, it is evident that Nb$_3$C$_2$O$_2$ exhibits the highest $\omega_{log}$, reaching 272meV, followed by  Nb$_4$C$_3$O$_2$ with a $\omega_{log}$ of 243meV. The discrepancy in $\omega_{log}$ between Nb$_3$C$_2$ and Ta$_3$C$_2$O$_2$ is relatively narrow, with values of 113meV and 129meV, respectively. This observation suggests that the reduction in T$_c$ during the transition from Nb$_3$C$_2$O$_2$ to Nb$_4$C$_3$O$_2$ primarily stems from the diminished $\omega_{log}$. Despite a minor increase in $\lambda$ during the transition, it fails to adequately counterbalance the effects of the reduced $\omega_{log}$, resulting in a decline in T$_c$. Conversely, the modest rise in temperature from Nb$_3$C$_2$O$_2$ to Ta$_3$C$_2$O$_2$ can be attributed to a significant drop in $\omega_{log}$, which is offset by the contributions from $\lambda$, ultimately leading to an overall upward trend in temperature. 

\begin{figure}[tb!]
\centerline{\includegraphics[width=0.5\textwidth]{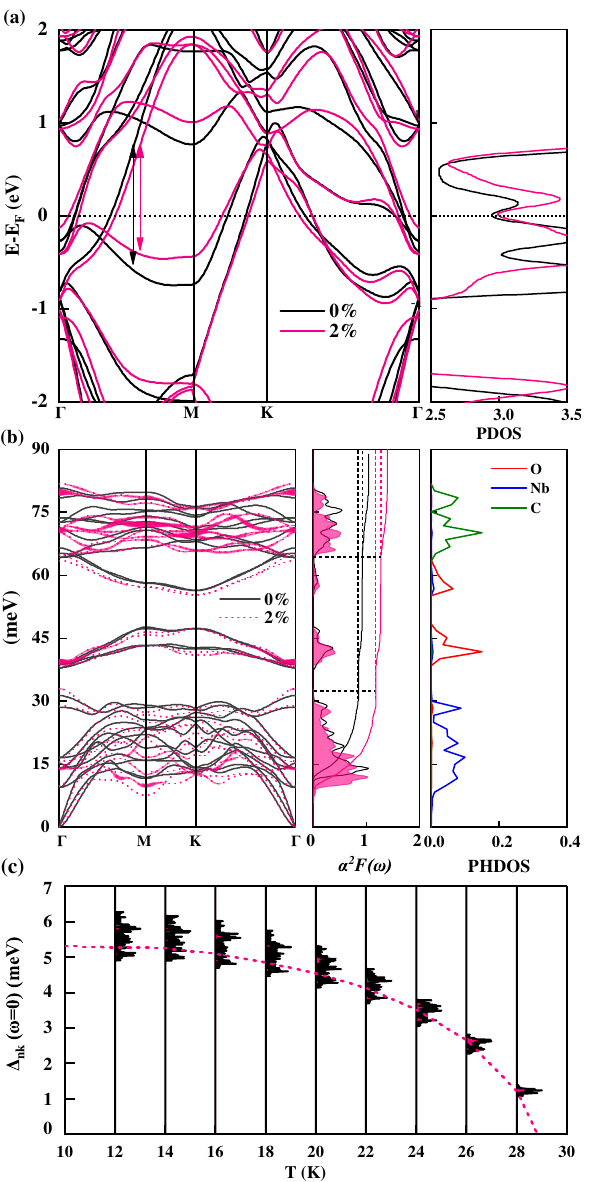}}
\caption{(a)Band structure, density of states (DOS). (b)Phonon spectrum, Eliashberg spectral function (${\alpha}^2F(\omega)$), electron-phonon coupling constant ($\lambda$), phonon density of states (PHDOS). (c) Superconducting transition temperature T$_c$ of Nb$_4$C$_3$O$_2$ at 2${\%}$ strain.
\label{fig:654-E-V}}
\end{figure}

\begin{figure}[tb!]
\centerline{\includegraphics[width=0.5\textwidth]{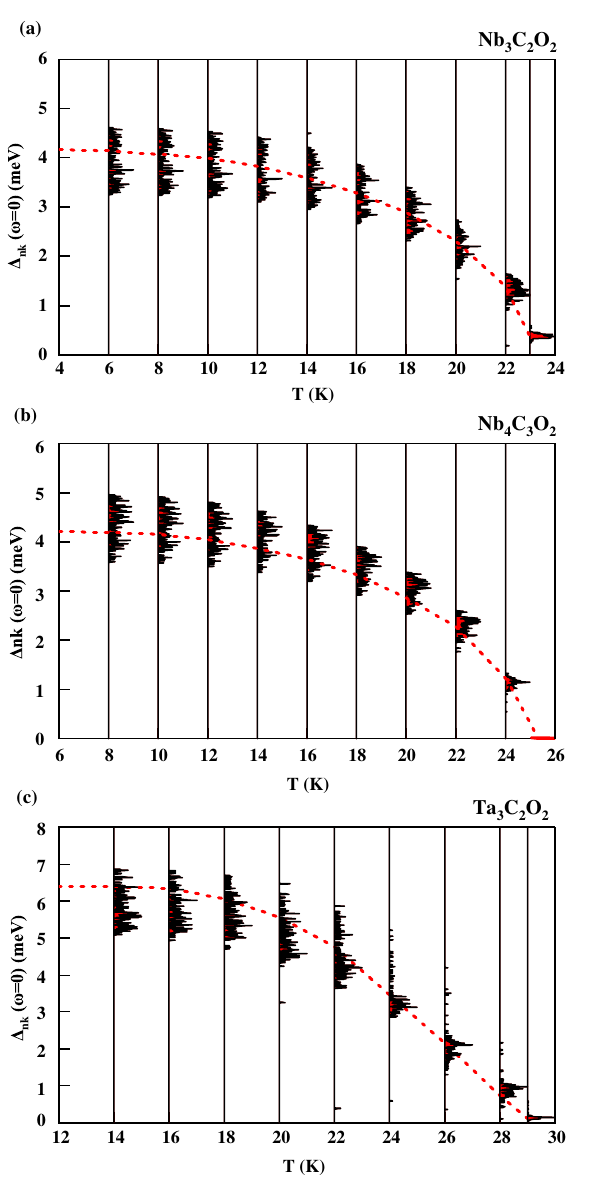}}
\caption{(a)(b)(c) Superconducting transition temperature T$_c$ of MXenes Nb$_3$C$_2$O$_2$, Nb$_4$C$_3$O$_2$ and Ta$_3$C$_2$O$_2$.
\label{fig:64-E-V}}
\end{figure}

The impact of strain on the superconducting properties of materials has been supported by various studies. According to Table S1, which outlines various parameters under strains ranging from 0${\%}$ to 4${\%}$, Nb$_4$C$_3$O$_2$ demonstrates the highest superconducting transition temperature (T$_c$) under a 2${\%}$ strain. To comprehend the mechanisms behind this enhancement in T$_c$, we delve into the electronic structure, structural stability, and behavior under 2${\%}$ strain. In Figure 3(a), when comparing the band structures of Nb$_4$C$_3$O$_2$ under 2${\%}$ strain with its unstrained counterpart, we can discover a noticeable band contraction at the Fermi level during strained situations. This contraction leads to an increased density of states (DOS) at the Fermi level, thereby positively impacting Nb$_4$C$_3$O$_2$'s superconducting performance under 2${\%}$ strain. Through further inspection of the phonon spectra in Figure 3(a), significant softening of the acoustic branches (highlighted by pink dashed lines) at the M and K points can be discovered. This softening precedes the appearance of the Eliashberg spectral function ${\alpha}^2F(\omega)$, which leads to an enhanced EPC strength $\lambda$, as depicted in Figure 3(b). A detailed analysis of $\lambda$ variation under strain reveals notable increases in contributions from all atoms: the $\lambda$ value attributed to Nb atoms rises by 38.37${\%}$, from 0.86 to 1.19, while the contributions from O and C atoms ascend from 0.07 and 0.11 to 0.09 and 0.12, respectively. This comprehensive augmentation indicates that under 2${\%}$ strain, each atom in Nb$_4$C$_3$O$_2$ contributes to boosting $\lambda$, with Nb atoms playing a prominent role. These collective contributions result in a $^{\sim}35$${\%}$ increase in the total EPC strength $\lambda$. Application of the anisotropic Migdal-Eliashberg theory suggests that a 2${\%}$ strain would elevate Nb$_4$C$_3$O$_2$'s superconducting transition temperature from an unstrained 25.00K to 29.00K under strain, as depicted in Figure 3(c). This trend aligns with predictions from the McMillan-Allen-Dynes formula, which correlates temperature changes with variations in both $\omega_{log}$ and $\lambda$. The observed increase in $\lambda$ has led to changes in $\omega_{log}$, resulting in a shift in T$_c$. The $\omega_{log}$ values for unstrained and 2${\%}$ strained Nb$_4$C$_3$O$_2$ at 243meV and 206meV are listed in TABLE S1.Under 2${\%}$ strain, the decrease in $\omega_{log}$ does not contribute to the T$_c$ enhancement but rather moderates the positive effect of the increased $\lambda$, leading to a modest T$_c$ increase from 19.49K to 24.80K.

Figure 4 illustrates the relationship between the superconducting energy gap ($\Delta_k$) and temperature based on the anisotropic Migdal-Eliashberg theory. It distinctly shows that M$_{\mathrm{n}+1}$C$_\mathrm{n}$O$_2$ (M=Nb,Ta) exhibits a single superconducting gap. For instance, at 6K as in Figure 4(a), there is a solitary gap ranging from 3.20meV to 4.80meV. This gap gradually diminishes and shifts downwards as the temperature increases, ultimately vanishing at 23.00K, indicating the critical temperature T$_c$. Consequently, the T$_c$ values for M$_{\mathrm{n}+1}$C$_\mathrm{n}$O$_2$ (M=Nb,Ta) are 23.00K, 25.00K, and 29.00K, respectively, all of which are slightly higher than the values derived from the McMillan-Allen-Dynes formula.

To conclude, this study delves into the MXenes M$_{\mathrm{n}+1}$C$_\mathrm{n}$O$_2$ (M=Nb,Ta), with a focus on electronic structure, structural and stability characteristics, and the superconducting transition temperature T$_c$. The findings we observed align with the McMillan-Allen-Dynes formula, emphasizing the intricate relationship between T$_c$, the logarithmic average of the phonon frequency ($\omega_{log}$), and the EPC constant ($\lambda$). Notably, the increase in $\lambda$ is primarily attributed to the softening of the transverse acoustic modes, a phenomenon driven by contributions from O atoms and M (M = Nb, Ta) atoms. This investigation highlights the exceptional superconducting properties of MXenes M$_{\mathrm{n}+1}$C$_\mathrm{n}$O$_2$ (M=Nb,Ta), positioning them as a valuable material for exploring 2D superconductivity and advancing the development of innovative superconducting devices.

\section{\label{sec:level1} ACKNOWLEDGMENTS}

This work is supported by National Natural Science Foundation of China (11904312, 11904313), Innovation Capability Improvement Project of Hebei province (22567605H), the Hebei Natural Science Foundation (A2022203006), Science Research Project of Hebei Education Department (BJK2022002).

\end{document}